\documentclass[conference]{IEEEtran}
\IEEEoverridecommandlockouts
\usepackage{cite}
\usepackage{amsmath,amssymb,amsfonts}
\usepackage{algorithm}
\usepackage{algpseudocode}
\usepackage{graphicx}
\usepackage{textcomp}
\usepackage{xcolor}
\usepackage{color}
\usepackage{colortbl}

\usepackage{textcomp}
\usepackage{stfloats}
\usepackage{url}
\usepackage{verbatim}

\usepackage[colorlinks, citecolor=black]{hyperref}

\usepackage{mathrsfs}
\usepackage{diagbox}

\usepackage{makecell}
\usepackage{cellspace}

\usepackage{longtable}
\usepackage{booktabs}
\usepackage{multirow,multicol}
\usepackage{float}

\usepackage[font=footnotesize]{caption}
\usepackage{subcaption}
\usepackage{lipsum}
\definecolor{mygray}{gray}{0.6}
\definecolor{myblue}{rgb}{0.8,0.85,1}

\usepackage{array, tabularx, boldline}
\usepackage{eurosym}
\usepackage{amstext} 
\usepackage{tabularray}

\def\BibTeX{{\rm B\kern-.05em{\sc i\kern-.025em b}\kern-.08em
    T\kern-.1667em\lower.7ex\hbox{E}\kern-.125emX}}
\begin{document}

\title{Efficient Privacy-Preserving Retrieval Augmented Generation with Distance-Preserving Encryption}

\author{
	\IEEEauthorblockN{Huanyi Ye\IEEEauthorrefmark{1}, Jiale Guo\IEEEauthorrefmark{2}, Ziyao Liu\IEEEauthorrefmark{2}, Kwok-Yan Lam\IEEEauthorrefmark{1}\IEEEauthorrefmark{2}}
\IEEEauthorblockA{\IEEEauthorrefmark{1}\textit{College of Computing and Data Science}, 
\textit{Nanyang Technological University}, 
Singapore \\
huanyi001@e.ntu.edu.sg}
\IEEEauthorblockA{\IEEEauthorrefmark{2}\textit{Digital Trust Centre (DTC)}, 
	Singapore  \\
	\{jiale.guo, liuziyao, kwokyan.lam\}@ntu.edu.sg }

}

\maketitle

\begin{abstract}
Retrieval-Augmented Generation (RAG) has emerged as a key technique for enhancing response quality of large language models (LLMs) without incurring high computational cost. It works by retrieving knowledge and fact from external databases, augmenting the prompt with the retrieved data, and enabling the LLM to generate more accurate responses. In traditional architectures, RAG services are provided by a single entity that hosts the dataset and issues queries within a trusted local environment. However, individuals or small organizations often lack the resources to maintain data storage servers, leading them to rely on outsourced, cloud-based storage for scalability. This dependence on untrusted third-party services introduces significant privacy risks, particularly when handling private and sensitive data. Embedding-based retrieval mechanisms, commonly used in RAG systems, are especially vulnerable to privacy leakage such as vector-to-text reconstruction attacks and structural leakage via vector analysis. Several privacy-preserving RAG techniques have been proposed to address such vulnerabilities; however, most existing approaches rely on partially homomorphic encryption, which incurs substantial computational overhead and limits their practicality in real-world deployments.\\
\hspace*{1em} To address these challenges, we propose an efficient privacy-preserving RAG framework (ppRAG) tailored for untrusted cloud environments that defends against vector-to-text attack, vector analysis, and query analysis. At its core, we propose Conditional Approximate Distance-Comparison-Preserving Symmetric Encryption (\texttt{CAPRISE}) that encrypts embeddings while still allowing the cloud to compute similarity between an encrypted query embedding and the encrypted database embeddings. \texttt{CAPRISE} preserves only the relative distance ordering between the encrypted query and each encrypted database embedding, without exposing inter-database distances, thereby enhancing both privacy and efficiency. To further mitigate query analysis risks, we introduce differential privacy by perturbing the query embedding prior to encryption, preventing the cloud from inferring sensitive patterns from query frequency. Experimental results show that ppRAG achieves efficient processing throughput, high retrieval accuracy, strong privacy guarantees, making it a practical solution for resource-constrained users seeking secure, cloud-augmented LLMs.
\end{abstract}

\begin{IEEEkeywords}
RAG, Privacy-Preserving Retrieval, Distance-Preserving Encryption, LLMs.
\end{IEEEkeywords}

\section{Introduction}
\label{sec:intro}

Large Language Models (LLMs), despite their powerful capabilities in generation and reasoning across various domains, face well-known limitations such as hallucination, outdated knowledge due to static training corpora, and lack of access to proprietary or time-sensitive information during inference\cite{brown2020language, ji2023survey}. To overcome these challenges, Retrieval-Augmented Generation (RAG) has emerged as a widely adopted framework. RAG enhances LLMs by coupling them with external retrieval databases: given a query, the model first retrieves relevant documents from a vectorized database, then generates responses based on the retrieved domain-specific content \cite{lewis2020retrieval, izacard2021leveraging}. 

RAG has seen significant industrial adoption, enabling not only large enterprises but also small organizations and individual users to deploy personalized LLM-based assistants at low cost. For example, on-device RAG deployments, such as those on smartphones or edge devices, can directly access personal data without relying on constant cloud connectivity, mitigating some privacy concerns. However, due to limited device storage and computing, users often outsource large volumes of data to untrusted cloud storage. This setup introduces a critical challenge: while encryption can protect data at rest, it complicates the retrieval process during query execution. Without a well-designed retrieval mechanism, users must download and decrypt large datasets locally, incurring significant bandwidth and computational overhead. A similar dilemma arises for small-to-medium enterprises that rely on outsourced, cloud-based storage to manage proprietary knowledge. Although this reduces device costs, it amplifies the privacy and efficiency challenges when sensitive data must be retrieved securely.

To enable document retrieval, RAG systems rely on efficient similarity search over stored content. Typically, documents are first embedded and stored in a vector database, and incoming queries are embedded in the same space to retrieve the top-$k$ most relevant results. To protect data privacy, these embedding vectors should also be stored using searchable encryption techniques, as storing them in plaintext posts substantial privacy risks~\cite{curtmola2006searchable}. However, most existing searchable encryption methods are ill-suited for the high-dimensional similarity computations that RAG systems demand. Recent research has explored the use of homomorphic encryption (HE) to protect embeddings~\cite{cheng2024remoterag}. While HE offers strong theoretical guarantees, its high computational cost makes it impractical for real-time RAG applications. This leads to a critical question: \textit{how can we support efficient similarity search over encrypted embeddings without leaking sensitive information}?

To address these challenges, we propose an efficient privacy-preserving framework for Retrieval-Augmented Generation (ppRAG). Our system enables users to encrypt their private embeddings and corresponding data before outsourcing them to the cloud, while still supporting efficient top-$k$ similarity search over encrypted embeddings without exposing their contents. At the core of ppRAG is a novel encryption scheme: Conditional Approximate Distance-Comparison-Preserving Symmetric Encryption (\texttt{CAPRISE}). \texttt{CAPRISE} allows the cloud server to compute relative distances between an encrypted query embedding and the encrypted database embeddings, crucial for retrieval, while ensuring that internal relationships between stored vectors remain obfuscated. This design explicitly defends against both vector-to-text reconstruction attacks and vector analysis.

To further protect against query-based inference attacks, where an attacker may infer sensitive information by analyzing repeated queries, ppRAG incorporates DistanceDP \cite{cheng2024remoterag}, a differentially private mechanism that perturbs the query embedding before encryption. This prevents the cloud from learning patterns in query frequency or similarity scores.

We formalize our design with provable privacy guarantees and validate its effectiveness through both theoretical analysis and empirical evaluation. Overall, ppRAG provides a practical, robust, and efficient solution for deploying secure RAG systems in untrusted cloud environments.

In summary, our contributions are as follows:
\begin{itemize}
    \item We introduce ppRAG, a practical framework for efficient privacy-preserving RAG that supports retrieval over outsourced encrypted databases.
    \item We design \texttt{CAPRISE}, a novel symmetric encryption scheme that preserves conditional distance comparison while obfuscating internal vector relationships, defending against Vec2Text and vector analysis attacks. In addition, we incorporate DP into the query phase to ensure differential privacy against repeated query analysis.
    \item We provide theoretical proofs, security analysis, and empirical evaluation demonstrating that ppRAG achieves privacy protection while maintaining efficiency.
\end{itemize}

\section{Related Work} \label{sec:related}

Retrieval-augmented generation (RAG) has been proposed to address the limitation of pre-trained models in updating or expanding their knowledge from external databases, a shortcoming that can lead to the generation of hallucinations \cite{lewis2020retrieval}. Widely research has been conducted in this field \cite{gao2023retrieval,wu2024retrieval,chen2024benchmarking,li2022survey}. Recent advancements have focused on improving retrieval mechanisms \cite{ma2023query,peng2024large,gao2023precise,sundfa,asaiself,wang2023self} and fine-tuning techniques \cite{wang2024unims,linra,wanginstructretro} to enhance the overall accuracy and relevance of generated prompts. 
Despite the rapid progress in RAG, few studies have focused on addressing the significant privacy challenges that arise when databases are outsourced to cloud platforms \cite{cheng2024remoterag} or when external data sources are integrated \cite{koga2024privacy}, potentially exposing sensitive information.

\renewcommand{\arraystretch}{1.2}
\begin{table}[th!]
    \centering
    \caption{Comparison of related work.}
    \label{tab:related}
    \begin{tabular}{c  c c c c c c} 
    \toprule    
        & Cloud & Local & DP & HE & OT & ADCPE \\
        \midrule
        \cite{ma2023query,peng2024large,gao2023precise,sundfa,asaiself,wang2023self,wang2024unims,linra,wanginstructretro} &  & $\checkmark$  &  &  &  &  \\ \hline
        \rowcolor{gray!10}RemoteRAG \cite{cheng2024remoterag} & $\checkmark$ &  & $\checkmark$ & $\checkmark$ & $\checkmark$ &  \\ \hline
        DP-RAG \cite{koga2024privacy} &  & $\checkmark$ & $\checkmark$ &  &  &  \\ \hline
        \rowcolor{gray!10}\textbf{ppRAG}* & $\checkmark$ &  & $\checkmark$ &  &  & $\checkmark$ \\
    \bottomrule
    \end{tabular}
\end{table}

RemoteRAG \cite{cheng2024remoterag} has been proposed to mitigate potential privacy leaks inherent in cloud-based RAG services. To safeguard sensitive query information from curious cloud providers, it introduces a technique called DistanceDP, which perturbs the embeddings before they are sent to the cloud, an important method given that embeddings can potentially be reconstructed back into their original text \cite{morris2023text}. Additionally, RemoteRAG employs partially homomorphic encryption (PHE) to compute distances between the encrypted query embeddings and stored documents, facilitating the retrieval of the top-$k$ results without exposing raw data. The system further incorporates oblivious transfer (OT) protocols to handle scenarios where privacy leakage is detected. However, the documents in RemoteRAG are stored in plaintext, and the computational overhead associated with PHE remains a significant challenge.

DP-RAG \cite{koga2024privacy} addresses concerns that a RAG system's generated output might leak sensitive information from a sensitive corpus, and thus proposes the use of DP to protect data. This approach assumes that the external database is fully integrated with both the retriever and generator. However, it does not consider scenarios where the data holder acts as the cloud provider, a critical omission given that maintaining a database can be challenging for small organizations and individuals, especially as data volumes grow and on-device LLMs become increasingly popular.

In light of these limitations, our work addresses the challenges encountered by existing privacy-preserving RAG methods, as shown in Table \ref{tab:related}. We introduce a novel approach, ppRAG, which offers computational accuracy and efficiency design. Our solution effectively tackles issues related to secure data storage and data trust, providing a more robust framework for privacy-preserving retrieval-augmented generation.

\begin{figure*}[th!]
    \centering
    \includegraphics[width=.7\linewidth]{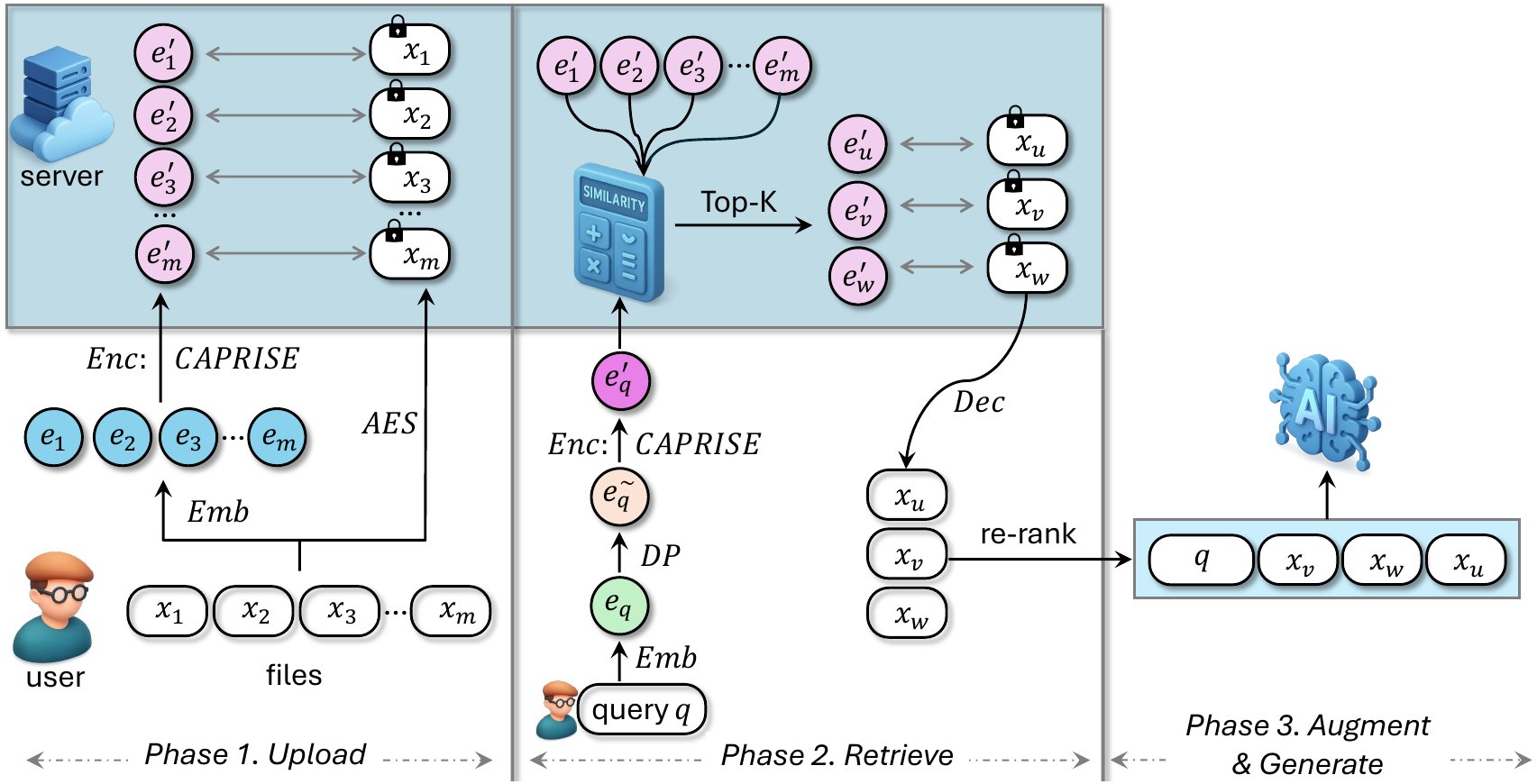}
    \caption{Overview of \textit{ppRAG}. In Phase 1 (Upload), the user embeds and encrypts local files, sending both AES-encrypted content and   \texttt{CAPRISE}-encrypted embeddings to the untrusted cloud. In Phase 2 (Retrieve), the user encrypts the query embedding and retrieves Top-K similar encrypted vectors from the server, which are then decrypted locally. In Phase 3 (Augment \& Generate), the retrieved content is combined with the original query to enhance LLM-based response generation.}
    \label{fig:overview}
\end{figure*}

\section{Methodology} 
\label{sec:methodology}

In this section, we detail the threat model and the intuition of our proposed method. Then we present our proposed architecture \textit{ppRAG} along with its key design steps for defending against vector-to-text attacks, vector analysis attacks, and query analysis attacks.

\subsection{Threat Model and Design Intuition}

We consider a scenario involving two main participants: the user and the cloud server. (i) The User: The user has limited local storage and must outsource private data (e.g., personal files) to the cloud. To enable retrieval-augmented generation (RAG), the user embeds these files and uploads both the encrypted embeddings and the corresponding AES-encrypted files to the cloud. At query time, the user embeds the query, sends the embedding to the cloud, and receives the top-k most similar encrypted files. These files are then decrypted locally and used to augment the query before being passed to a large language model (LLM) for inference. (ii) The Cloud Server: The server stores the uploaded embeddings and encrypted files. Upon receiving a query embedding, it performs similarity search and returns the top-k nearest neighbors based on the distance between the query and stored embeddings.

In our threat model, the cloud server is assumed to be honest-but-curious: it performs all expected operations correctly but is not trusted with the user’s private data and may attempt to infer sensitive information.

This setting introduces multiple privacy risks. First, storing raw embedding vectors in the cloud is inherently unsafe. Prior work~\cite{morris2023text, zhuang2024understanding} has shown that embeddings can be inverted to recover the original text data with high fidelity, rendering AES encryption ineffective if embeddings are exposed. Second, by analyzing distances between stored embeddings, the server can infer semantic relationships between the user's files, even when no queries are made. Third, through repeated user queries, the server can identify frequently accessed records, revealing usage patterns.

To mitigate these risks, we assume that the user performs local embedding and encryption before uploading data to the cloud. Our design goal is to protect both the semantic content of the embeddings and the relational information between them, even under repeated access patterns.

\subsection{Proposed Architecture ppRAG}

Our proposed architecture, ppRAG, is designed for scenarios where a user needs to outsource a private document database to an untrusted third party, such as the cloud; an overview of the entire process is shown in Figure~\ref{fig:overview}. Since the cloud server is not trusted, the user must encrypt the files $\{x_1, x_2, \ldots, x_m\} $ before uploading in Phase 1. However, RAG systems require comparing the similarity between a user query and the stored documents, which in turn depends on preserving the distance relationships between their representations. To enable similarity-based retrieval while ensuring privacy, the user first embeds each file to get $\{e_1, e_2, \ldots, e_m\} $ and generates corresponding AES-encrypted versions $\{\mathbf{Enc}(x_1), \mathbf{Enc}(x_2), \ldots, \mathbf{Enc}(x_m)\} $ of the original content. These embeddings $\{e_1, e_2, \ldots, e_m\} $ are then paired with their encrypted files  $\{\mathbf{Enc}(x_1), \mathbf{Enc}(x_2), \ldots, \mathbf{Enc}(x_m)\} $ and uploaded to the cloud. The cloud uses the embeddings to perform similarity search and returns the top-k encrypted files to the user, who decrypts them locally for query augmentation and LLM-based generation.

Given the privacy risks posed by the potential inversion of embedding vectors ${e_1, e_2, \ldots, e_m}$ to recover the original text, a threat known as the \textit{Vec2Text Attack}\cite{morris2023text}, we initially adopt the Approximate-Distance-Comparison-Preserving Encryption scheme, \texttt{ADCPE}\cite{fuchsbauer2022approximate}, which allows similarity search while protecting the raw embeddings. 

However, preserving all pairwise distances, as \texttt{ADCPE} does, may unnecessarily expose relational information among stored embeddings. To mitigate this \textit{Vector Analysis Attack}, we propose Conditional ADCPE (\texttt{CAPRISE}), an efficient encryption scheme that preserves only the distances between the query embedding $e_q$ and each encrypted database embedding $\{e'_1, e'_2, \ldots, e'_m\}$, which is sufficient for retrieval, while obfuscating inter-database distances to prevent structural analysis. We use letters with a ``prime" symbol to represent the encrypted embeddings, e.g., $e'_1$ is the ciphertext corresponding to the plaintext embedding $e_1$.

In Phase 2, even if the query embedding $e_q$ is encrypted using \texttt{CAPRISE}, it may still leak information under repeated queries. For example, a curious server could infer sensitive patterns by analyzing query frequencies or similarity scores. To prevent this \textit{Query Analysis Attack}, we apply differential privacy to perturb the query embedding before transmission, resulting in a noised version $\tilde{e}_q$. To ensure compatibility with the encrypted database embeddings $\{e'_1, e'_2, \ldots, e'_m\}$, the perturbed query embedding $\tilde{e}_q$ is further encrypted using \texttt{CAPRISE} , denoted as $e'_q$, to preserve distance relationships needed for similarity computation.

The following subsections detail our defenses against (i) \textit{Vec2Text Attacks} in Phase 1, (ii) \textit{Vector Analysis} in the encrypted database, and (iii) \textit{Query Analysis} attacks in Phase 2.

\subsubsection{\textbf{Defense against Vec2Text Attacks and Vectors Analysis}} \label{sec:defense_vec2text}
As explained above, directly uploading and storing only the encrypted files $\{\mathbf{Enc}(x_1), \mathbf{Enc}(x_2), \ldots, \mathbf{Enc}(x_m)\}$ on the cloud server does not ensure correct retrieval in a RAG system. To support similarity-based retrieval, our framework requires the use of embeddings as indices corresponding to the encrypted files. These embeddings are used to compute similarity during retrieval.

However, embeddings can potentially be inverted to recover the original text \cite{morris2023text}, posing a significant privacy risk. Thus, it is essential to ensure that distance comparisons are preserved while also protecting the embeddings. To achieve this, we initially consider the Approximate Distance Comparison Preserving Encryption (\texttt{ADCPE}) method~\cite{fuchsbauer2022approximate}, which preserves relative distance between plaintext vectors after encryption. Specifically, given three vectors, $e_1$ $e_2$ and $e_3$, for any $\mathbf{s} \in \mathcal{S}$ where $\mathcal{S}$ is the keyspace, and $\beta$ is to satisfy the \textit{ distance-preserving}: 
\begin{align}
\label{eq:ADCPE}
&\quad \|e_1 - e_2\| < \|e_1 - e_3\| - \beta \Rightarrow 
\|e'_1 - e'_2\| < \|e'_1 - e'_3\| \notag\\
&where \ \
e'_1 = \mathbf{Enc}(s, e_1), \, e'_2 = \mathbf{Enc}(s, e_2), \,e'_3 = \mathbf{Enc}(s, e_3)\
\end{align}

\begin{algorithm}[tbp!] 
\caption{Conditional Approximate Distance-Comparison-Preserving Symmetric Encryption (\texttt{CAPRISE})}
\label{alg:CAPRISE}
\begin{algorithmic}[1]

\Procedure{KeyGen}{}
    \State $s \xleftarrow{\$} \mathcal{S}$
    \State $K \xleftarrow{\$} \{0,1\}^k$
    \State \textbf{return} $(s, K)$
\EndProcedure

\vspace{0.2cm}

\Procedure{Enc$_{\text{db}}$}{$(s, K), e_i$}
    \State $r \xleftarrow{\$} \{0,1\}^\ell$
    \State $\text{randbits}_1 \| \text{randbits}_2 \gets \text{PRF}(K, r)$
    \State $\mathbf{n} \gets \mathcal{N}_{(0,I_d)}(\text{randbits}_1)$
    \State $u \gets \mathcal{U}_{0,1}(\text{randbits}_2)$
    \State $\lambda_{e_i} \gets \frac{3} {8} \frac{\mathbf{n} s \beta}{\|\mathbf{n}\|}(u)^{\frac{1}{d}}$
    \State $e'_i \gets s e_i + \lambda_{e_i}$
    \State \textbf{return} $(e'_i, r)$
\EndProcedure

\vspace{0.2cm}

\Procedure{Dec$_{\text{db}}$}{$(s, K), (e'_i, r)$}
    \State $\text{randbits}_1 \| \text{randbits}_2 \gets \text{PRF}(K, r)$
    \State $\mathbf{n} \gets \mathcal{N}_{(0,I_d)}(\text{randbits}_1)$
    \State $u \gets \mathcal{U}_{0,1}(\text{randbits}_2)$
    \State $\lambda_{e_i} \gets \frac{3} {8} \frac{\mathbf{n} s \beta}{\|\mathbf{n}\|}(u)^{\frac{1}{d}}$
    \State $e_i \gets \frac{e'_i - \lambda_{e_i}}{s}$
    \State \textbf{return} $e_i$
\EndProcedure

\vspace{0.2cm}

\Procedure{Enc$_{\text{q}}$}{$(s, K), e_q$}
    \State $r \xleftarrow{\$} \{0,1\}^\ell$
    \State $\text{randbits}_1 \| \text{randbits}_2 \gets \text{PRF}(K, r)$
    \State $\mathbf{n} \gets \mathcal{N}_{(0,I_d)}(\text{randbits}_1)$
    \State $u \gets \mathcal{U}_{0,1}(\text{randbits}_2)$
    \State $\eta_{e_q} \gets \frac{1} {8} \frac{\mathbf{n} s \beta}{\|\mathbf{n}\|}(u)^{\frac{1}{d}}$
    \State $e'_q \gets s e_q + \eta_{e_q}$
    \State \textbf{return} $(e'_q, r)$
\EndProcedure

\end{algorithmic}
\end{algorithm}

\begin{figure}[b]
    \centering
    \includegraphics[width=.8\linewidth]{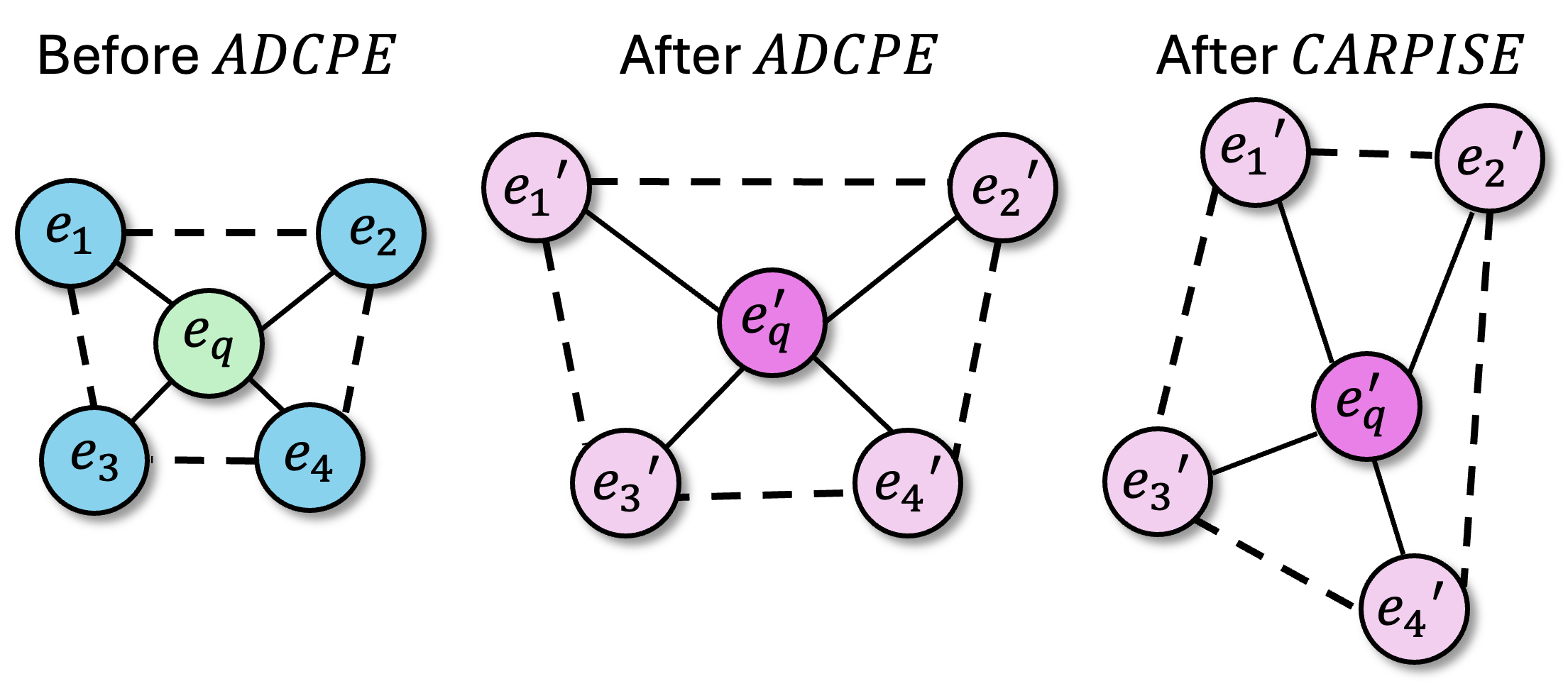}
    \caption{Illustration of \texttt{ADCPE} and \texttt{CAPRISE}. While the original embedding vectors $\{e_1, e_2, e_3, e_4\}$ and query embedding $e_q$ are encrypted into $\{e'_1, e'_2, e'_3, e'_4\}$ and $e'_q$, the ranking of distances is preserved under \texttt{ADCPE}. As illustrated, solid edges represent distances between the query and database embeddings, and dashed edges represent distances among database embeddings. \texttt{ADCPE} maintains both, potentially leaking global relational structure. \texttt{CAPRISE} mitigates this by preserving only the query-to-database (solid) distances while perturbing inter-database (dashed) distances, thereby protecting internal relationships in the encrypted space.}
    \label{fig:vector_analysis}
\end{figure}

While \texttt{ADCPE} effectively mitigates Vec2Text attacks and other vulnerabilities discussed above, it still introduces additional privacy concerns because the encrypted embeddings stored in the cloud database continue to expose relative distance information, which may reveal structural relationships among the embeddings. To mitigate this issue, we propose an enhanced and efficient \texttt{ADCPE} scheme that prevents the cloud server from analyzing distances among embeddings stored in the database. We name this scheme \textit{Conditional Approximate Distance-Comparison-Preserving Symmetric Encryption} (\texttt{CAPRISE}). Unlike standard \texttt{ADCPE}, \texttt{CAPRISE} is \textit{conditional} in the sense that it does not preserve distances between encrypted embeddings $\{e'_1, e'_2, \ldots, e'_m\}$ stored in the cloud. Instead, it solely preserves distance comparisons only between the encrypted query embedding $e'_q$, submitted by the user in Phase 2, and each encrypted database embedding $e'_i \in \{e'_1, e'_2, \ldots, e'_m\}$, as illustrated in Figure~\ref{fig:vector_analysis}. This ensures that similarity-based retrieval remains functional, while relational structure within the dataset is obfuscated to protect against vector analysis attacks.

To be more specific, \texttt{CAPRISE} introduces two encryption algorithms: (i) $\mathbf{Enc_{\text{DB}}}$, used to encrypt embeddings stored on the cloud server in Phase 1, and (ii) $\mathbf{Enc_{\text{Q}}}$, used to encrypt query embeddings in Phase 2. Embeddings encrypted with $\mathbf{Enc_{\text{DB}}}$ do not retain meaningful distance relationships among themselves, thereby mitigating structural leakage. However, \texttt{CAPRISE} ensures that distance relationships between the encrypted query embedding and the encrypted database embeddings, i.e., between $\mathbf{Enc_{\text{Q}}}(e_q)$ and each $\mathbf{Enc_{\text{DB}}}(e_i)$, are preserved, allowing for accurate similarity-based retrieval without compromising internal database structure. The procedures are detailed in Algorithm~\ref{alg:CAPRISE}, where $\mathcal{S}$ denotes the key space with $|\mathcal{S}| \leq 2\lambda$, and $\lambda$ is the security parameter. The function $\mathsf{PRF}: \{0,1\}^k \times \{0,1\}^\ell \rightarrow \{0,1\}^*$ denotes a pseudorandom function family for some $k, \ell \in \mathbb{N}$, with $K$ and $r$ as its inputs. The symbols $\mathbf{n}$ and $u$ refer to samples drawn from the standard Gaussian and uniform distributions, respectively. In addition to $\mathbf{Enc_{\text{DB}}}$ and $\mathbf{Enc_{\text{Q}}}$, Algorithm~\ref{alg:CAPRISE} also includes the procedures for $\mathsf{KeyGen}$, which generates the key pair $(s, K)$, and $\mathbf{Dec_{\text{DB}}}$, which decrypts the embeddings retrieved from the database.

Formally, \texttt{CAPRISE} is required to satisfy the following property:

\textbf{Claim:} Given database embeddings $e_1$, $e_2$, $e_3$, and a query embedding $e_q$, for any $\mathbf{s} \in \mathcal{S}$, the following holds:

\begin{align}
&\, \|e_q - e_1\| < \|e_q - e_2\| - \beta \Rightarrow \|e'_q - e'_1\| < \|e'_q - e'_2\|  \label{eq:CAPRISE_1} \\
&\, \|e_1 - e_2\| < \|e_1 - e_3\| - \beta \nRightarrow \|e'_1 - e'_2\| < \|e'_1 - e'_3\| \label{eq:CAPRISE_2}\\
& \quad where \quad e'_q = \mathbf{Enc_{\text{Q}}(s, e_q)}, \, e'_1 = \mathbf{Enc_{\text{DB}}(s, e_1)}, \notag \\
& \,\,\,\, \qquad \qquad e'_2 = \mathbf{Enc_{\text{DB}}(s, e_2)}, \, e'_3 = \mathbf{Enc_{\text{DB}}(s, e_3)} \notag
\end{align}

Equation~\eqref{eq:CAPRISE_1}~\eqref{eq:CAPRISE_2} is achieved by carefully modifying the noise component of the encryption algorithm. The proof is provided below.

\noindent \textit{Proof.} Let $e_1, e_2, \ldots,e_m, e_q \in \mathcal{M}$ be embeddings in the plaintext space. Assume the following inequalities hold:
\begin{align}
\label{eq:embedding_inequality}
&\|e_q - e_1\| < \|e_q - e_2\| - \beta \notag \\ 
\text{and} \quad &\|e_1 - e_2\| < \|e_1 - e_3\| - \beta.
\end{align}

For notational simplicity, we denote the encryption functions as $\mathcal{C}_{\text{Q}}(\cdot) = \mathbf{Enc_{\text{Q}}}(s, \cdot)$ for query embeddings and $\mathcal{C}_{\text{DB}}(\cdot) = \mathbf{Enc_{\text{DB}}}(s, \cdot)$ for database embeddings. These encryptions incorporate noise designed to balance privacy and retrieval utility. Specifically, we define:
\begin{align}
&e'_q = \mathcal{C}_{\text{Q}}(e_q) = se_q + \eta_{e_q} \label{eq:embedding_encryption_eq}\\
&e'_i = \mathcal{C}_{\text{DB}}(e_i) = se_i + \lambda_{e_i} \ (i\in\{1,2,\ldots,n\}) \label{eq:embedding_encryption_ei}
\end{align}
where $\eta_{e_q}$ and $\lambda_{e_i}$ are the noise vectors added to the query and database embeddings, respectively, satisfying the bounds we defined in \texttt{CAPRISE}:
\begin{align}
\label{eq:bound_eta_lambda}
\|\eta_{e_q}\| < \frac{s\beta}{8}, \quad \|\lambda_{e_i}\| < \frac{3s\beta}{8}
\end{align}

To prove the \textit{distance-comparison-preserving} property between the encrypted query embedding $e'_q$ and database embedding $e'_i$, as described in Inequality~\eqref{eq:CAPRISE_1}, we begin with $\|e_q - e_1\| < \|e_q - e_2\| - \beta$, and show that this ordering is preserved after encryption, i.e., $\|e'_q - e'_1\| < \|e'_q - e'_2\|$. The proof proceeds as follows.
\begin{align}
\|e'_q - e'_1\| &= \|\mathcal{C}_{\text{Q}}(e_q) - \mathcal{C}_{\text{DB}}(e_1)\| \notag \\
&\leq \|\mathcal{C}_{\text{Q}}(e_q) - se_q\| + \|se_q - se_1\| + \|\mathcal{C}_{\text{Q}}(e_q) - se_1\| \label{eq:proof_triangle}\\
&= \|\eta_{e_q}\| + s\|e_q - e_1\| + \|\lambda_{e_1}\|  \notag \\
&< \|\eta_{e_q}\| + \|\lambda_{e_1}\| + s(\|e_q - e_2\| - \beta) \label{eq:proof_assum1} \\
&< \frac{s\beta}{8} + \frac{3s\beta}{8} +\|se_q - se_2\| - s\beta \label{eq:proof_assum2}\\
&= \|se_q - se_2\| - \frac{s\beta}{2} \notag \\
&< \|se_q - se_2\| - (\|\eta_{e_q}\| + \|\lambda_{e_2}\|) \notag \\
&\leq \|se_q - se_2\| - \|\eta_{e_q} - \lambda_{e_2}\| \label{eq:proof_ti2}\\
&\leq \|(se_q - se_2) - (\eta_{e_q} - \lambda_{e_2})\| \label{eq:proof_ti3}\\
&= \|\mathcal{C}_{\text{Q}}(e_q) - \mathcal{C}_{\text{DB}}(e_2)\| \notag \\
&=\|e'_q - e'_2\| \notag
\end{align}

Inequality~\eqref{eq:proof_triangle} applies the triangle inequality, separating the distance $\|e'_q - e'_1\|$ into three components: the noise in the query encryption, the scaled distance between the unencrypted vectors, and the noise in the database encryption. 
Inequality~\eqref{eq:proof_assum1} substitutes the assumed ordering between $\|e_q - e_1\|$ and $\|e_q - e_2\|$  based on the assumption in Inequality~\eqref{eq:CAPRISE_1}.
Inequality~\eqref{eq:proof_assum2} follows from the noise bounds defined in \texttt{CAPRISE}, as specified in Inequalities~\eqref{eq:bound_eta_lambda}.
Inequalities~\eqref{eq:proof_ti2} and \eqref{eq:proof_ti3}) again use the triangle inequality in reverse to combine the scaled distance and noise difference into a single term, thereby bounding $\|e'_q - e'_2\|$.

Combining all these steps above shows that $\|e'_q - e'_1\| < \|e'_q - e'_2\|$, which completes the proof that the encryption preserves distance orderings relevant for retrieval.

To prove the Inequality~\eqref{eq:CAPRISE_2}, which states that the \textit{distance-comparison-preserving} property does \textit{not} hold  between encrypted database embeddings $e'_i$, we begin by upper-bounding the distance between the encrypted vectors $e'_1$ and $e'_2$:

\begin{align}
\|e'_1 - e'_2\| &= \|\mathcal{C}_{\text{DB}}(e_1) - \mathcal{C}_{\text{DB}}(e_2)\| \notag \\
&< \|\lambda_{e_1}\| + \|\lambda_{e_2}\| + s(\|e_1 - e_3\| - \beta) \label{eq:proof2_assum1} \\
&< \frac{3s\beta}{8} + \frac{3s\beta}{8} + \|se_1 - se_3\| - s\beta \label{eq:proof2_assum2} \\
&= \|se_1 - se_3\| - \frac{s\beta}{4} \notag
\end{align}

Inequalities~\eqref{eq:proof2_assum1} and~\eqref{eq:proof2_assum2} follow the same derivation steps as Inequalities~\eqref{eq:proof_assum1} and~\eqref{eq:proof_assum2} from the earlier proof. For simplicity, there must exist a small constant $\delta > 0$ such that:
\begin{equation}
\label{eq:proof_delta1}
\|e'_1 - e'_2\| = \|se_1 - se_3\| - \frac{s\beta}{4} -\delta 
\end{equation}

We now lower-bound the distance $\|e'_1 - e'_3\|$:
\begin{align}
\|e'_1 - e'_3\| &= \|\mathcal{C}_{\text{DB}}(e_1) - \mathcal{C}_{\text{DB}}(e_3)\| \notag\\
& = \|(se_1 - se_3) + (\lambda_{e_1} - \lambda_{e_3})\| \label{eq:db_equality}\\
&\geq \|se_1 - se_3\| - \|\lambda_{e_1} - \lambda_{e_3}\| \label{eq:db_triangle_inequality}\\
&> \|se_1 - se_3\| - (\|\lambda_{e_1}\| + \|\lambda_{e_3}\|) \label{eq:db_triangle_inequality_2}\\
&> \|se_1 - se_3\| -  \frac{3s\beta}{4} \label{eq:db_beta}
\end{align}

Equality~\eqref{eq:db_equality} follows from the definition provided in Equation~\eqref{eq:embedding_encryption_ei}.
Inequality~\eqref{eq:db_triangle_inequality} applies the reverse triangle inequality, while Inequality~\eqref{eq:db_triangle_inequality_2} applies the standard triangle inequality.
Finally, Inequality~\eqref{eq:db_beta} is derived from the noise bounds defined in~\eqref{eq:bound_eta_lambda}.

Again, there must exist a small constant $\delta' > 0$ such that:
\begin{equation}
\label{eq:proof_delta2}
\|e'_1 - e'_3\| = \|se_1 - se_3\| -  \frac{3s\beta}{4} + \delta' 
\end{equation}

If the combined deviation satisfies $\delta + \delta' < \frac{s\beta}{2}$, then comparing~\eqref{eq:proof_delta1} and~\eqref{eq:proof_delta2} gives:
\[\|e'_1 - e'_2\| > \|e'_1 - e'_3\|\] 

This result contradicts the original ordering of unencrypted distances $\|e_1 - e_2\| < \|e_1 - e_3\| - \beta$, thereby confirming that \texttt{CAPRISE} does \textit{not} preserve distance-based orderings among encrypted database embeddings. This intentional relaxation is essential to obscure relational structure in the stored embeddings and defend against vector analysis attacks.

\subsubsection{\textbf{Defense against Query Analysis}} \label{sec:defense_query}

\begin{figure}[b]
    \centering
    \includegraphics[width=.45\linewidth]{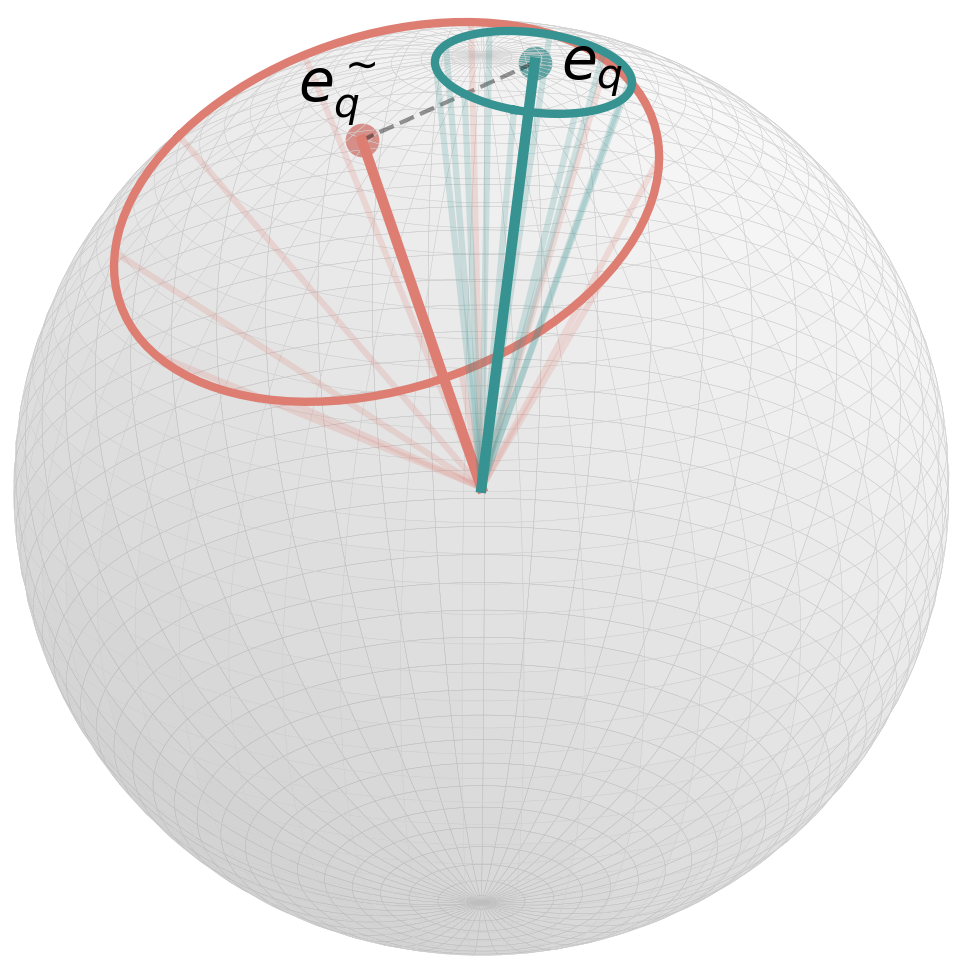}
    \caption{Illustration of query embedding perturbation by differential privacy. To protect user privacy from distance-based inference attacks across repeated queries, controlled noise is added to the query embedding, expanding the search region and preventing precise query analysis by the server.}
    \label{fig:sphere_projection}
\end{figure}

Repeated user queries can introduce additional privacy risks. A curious server may perform distance-based analysis across multiple queries, potentially inferring sensitive patterns such as query frequency such as the frequency with which certain database vectors are retrieved. To address this concern, we adopt \textit{DistanceDP}~\cite{cheng2024remoterag}, which is a differential privacy technique that obfuscates the query by adding calibrated noise to the query embedding $e_q$ and expanding the retrieval space, as illustrated in Figure~\ref{fig:sphere_projection}. This mechanism ensures that the server cannot reliably infer the true query content, thereby defending against query analysis attacks.

To formally support the effectiveness of DistanceDP, we quantify how much the retrieval set must expand to account for the angular distortion introduced by noise. The following theorem characterizes the relationship between the angular perturbation and the increase in top-$k'$ results needed to retain the original top-$k$ set. This provides a theoretical foundation for selecting an appropriate search radius.

\textbf{Theorem 1.} Let $e_q$ and $e^{\sim}_{q}$ be two query embeddings with a perturbed angular difference $\Delta \alpha_k$. To ensure that the top-$k'$ results retrieved using $e^{\sim}_{q}$ still include the top-$k$ results corresponding to $e_q$, the values $k'$ and $k$ must satisfy the following relationship:

\begin{equation}
\label{eq:dp_relationship}
    k'  = m \cdot \frac{\Omega_{n-1}(\pi)}{\Omega_n(\pi)} \cdot \int_{\alpha_k}^{\alpha_{k'}} \sin^{n-2} \theta \, d\theta \  + \  k
\end{equation}
where $\alpha_{k'} = \alpha_k + \Delta \alpha_k$ denotes the widened angular threshold due to noise injection, and $\Omega_n(\pi) = \frac{2\pi^{n/2}}{\Gamma(n/2)}$ represents the surface area of a unit $n$-dimensional hypersphere. This relationship quantifies how the search radius must expand to include the retrieval results despite the perturbation introduced for differential privacy, thereby supporting the robustness of our defense against query analysis attacks.

\begin{algorithm}[htbp!] 
\caption{Efficient Privacy-Preserving Retrieval-Augmented Generation (ppRAG)}
\label{alg:whole}
\begin{algorithmic}[1]
\Procedure{Phase 1: Data Preparation and Upload}{}
    \State \textbf{Input:} $\{x_1, x_2, \dots, x_m\}$
    \State $e_i \gets \mathsf{LLM_e}(x_i), \ i \in \{1, \dots, n\}$
    \State $(s, K) \gets \mathsf{KeyGen}$
    \State $e'_i \gets \mathbf{Enc_{\text{DB}}}\left( (s, K), e_i\right)$
    \State $k_{\text{AES}} \gets \mathsf{KeyGen_{\text{AES}}}$
    \State $x'_i \gets \mathbf{ENC}_{\text{AES}}(k_{\text{AES}}, x_i)$
    \State  \textbf{Upload} $\{(e'_1, x'_i), \dots, (e'_m, x'_m)\}$ to $\textsf{Server}$
\EndProcedure

\vspace{0.2cm}

\Procedure{Phase 2: Query and Retrieval}{}
    \State \textbf{Input:} query text $q$, $k$
    \State $e_q \gets \mathsf{LLM_e}(q)$
    \State $k',\ e^{\sim}_q \gets \mathsf{DistanceDP}(e_q)$
    \State $e'_q \gets \mathbf{Enc_{\text{Q}}}\left( (s, K), e^{\sim}_q \right)$
    \State Send $e_q'$ to $\textsf{Server}$ for retrieval
    \State \makebox[\linewidth][l]{$\{(e'_i, x'_i)\}_{i=1}^{k'} \gets \mathsf{Database\_Search}(e_q', k')$ {\scriptsize\Comment{Server-side}}}
    \State $\{e_i\}_{i=1}^{k'} \gets \mathbf{Dec_{\text{DB}}}\left((s, K), (e'_i, r) \right)$ 
    \State $\mathsf{Reranked}( \{e_j\}_{j=1}^{k'} ) \gets \mathsf{ReRank} \{e_i\}_{i=1}^{k'} $ 
    \State $\{(e_j, x_j)\}_{j=1}^{k} \gets \mathsf{Top_{k}}( \mathsf{Reranked}( \{e_j\}_{j=1}^{k'} )$
    \State \textbf{Output:} Retrieved query $\{x_1, x_2, \cdots, x_k\}$
\EndProcedure

\vspace{0.2cm}

\Procedure{Phase 3: Prompting the LLM}{}
    \State \textbf{Input:} $ q \, , \, \{x_1, x_2, \cdots, x_k\}$
    \State $ \mathsf{input} \gets q \, + \, \{x_1, x_2, \cdots, x_k\}$
    \State $\mathsf{Response} \gets \mathsf{LLM}(\mathsf{input})$
    \State \textbf{Output:} $\mathsf{Response}$
\EndProcedure

\end{algorithmic}
\end{algorithm}

The complete workflow of our proposed efficient ppRAG system is presented in the pseudocode of Algorithm~\ref{alg:whole}, which corresponds to the three-stage design illustrated in Figure~\ref{fig:overview}. In this algorithm, the $\mathsf{DistanceDP}$ mechanism, detailed in Subsection~\ref{sec:defense_query}, ensures differential privacy during query submission. The $\mathsf{Database\_Search}$ function retrieves the Top-$k'$ most similar encrypted embeddings to the query embedding $e'_q$ from the server-side database. To enhance result quality, $\mathsf{ReRank}$, a commonly used step in RAG systems, is applied on the retrieved embeddings to reorder them based on refined relevance scores. Finally, $\mathsf{Top}_k$ selects the top-$k$ most relevant encrypted data files from the reranked list to return to the user.

\section{Security Analysis} 
\label{sec:security_ana}

In this section, we prove that ppRAG is able to preserve user data privacy against the honest-but-curious adversary, especially the cloud server. It ensures that even with access to encrypted data and query embeddings, the server is unable to infer any meaningful information about the user's private inputs.

\subsubsection{CAPRISE Security Guarantee}
Our proposed encryption scheme \texttt{CAPRISE} inherits the security properties of \texttt{ADCPE}. The key enhancement in \texttt{CAPRISE}, namely, decoupling distance preservation between stored embeddings while maintaining it conditionally for query embeddings, does not compromise the original security guarantees. The encryption process introduces a carefully calibrated noise vector $\lambda_{e_i} \gets \frac{3}{8} \frac{\mathbf{n} s \beta}{|\mathbf{n}|}(u)^{1/d}$ into the embedding, such that it preserves the distance between the query embedding and each database embedding, while obfuscating the distances among the database embeddings themselves.

Such calibrated noise is crucial since modern Vec2Text attacks are highly sensitive to small changes in the input embeddings, as discussed in \cite{morris2023text}. Even minor perturbations can cause such attacks to reconstruct semantically unrelated or completely incorrect text. Our empirical results confirm that \texttt{CAPRISE} effectively defends against Vec2Text attacks by introducing noise sufficient to disrupt semantic meanings while still preserving retrieval accuracy.

\subsubsection{Query Privacy under DistanceDP}
The query embedding is randomized via DistanceDP mechanism, which adds DP noise and expands the retrieval space. This ensures that the server cannot reliably infer the original query or its semantic neighbors from the perturbed encrypted query vector.

Formally, DistanceDP satisfies $(\varepsilon, \delta)$-differential privacy with respect to the query embedding space. This obfuscation is sufficient to protect against query analysis attacks, such as those inferring repeated access patterns or semantically similar queries.

\section{Experimental Evaluation} \label{sec:experiments}

We conduct experiments on the MS MARCO dataset~\cite{nguyen2016ms}. For the embedding model, we adopt \texttt{gtr-t5-base}~\cite{ni2022large}. We evaluate \texttt{CAPRISE} from the perspectives of efficiency, privacy, and accuracy. All test experiments are conducted on an NVIDIA A100 GPU.

\subsubsection{Efficiency} \label{sec:exp_efficiency}
\begin{figure}[htbp!]
\centering
    \begin{subfigure}[b]{0.9\columnwidth}
    \includegraphics[width=\linewidth]{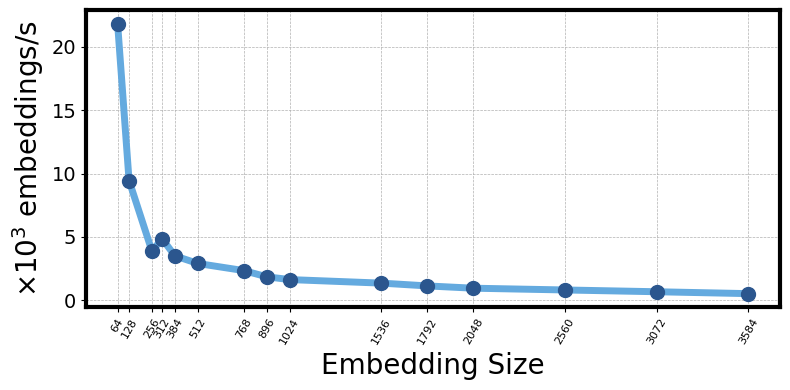}
    \caption{Encryption throughput of embeddings across varying embedding sizes}
    \label{fig:rate_processing}
    \end{subfigure}
\vspace{0.5em}
    \begin{subfigure}[b]{0.9\columnwidth}
    \includegraphics[width=\linewidth]{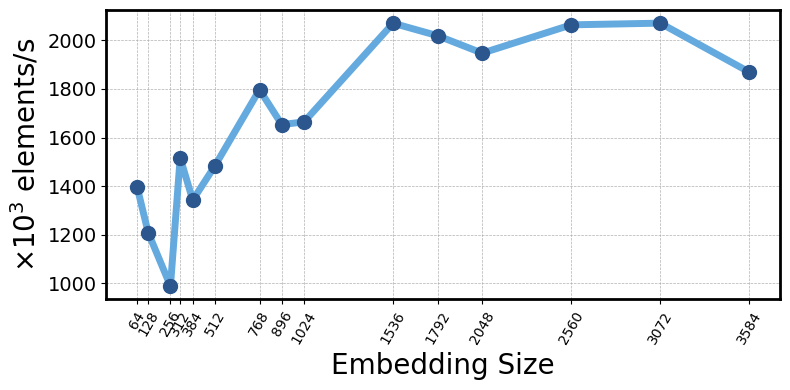}
    \caption{Encryption throughput of features across varying embedding sizes}
    \label{fig:rate_encryption}
    \end{subfigure}
\caption{Efficiency of $\mathsf{CAPRISE}$}
\label{fig:efficiency}
\end{figure}

We evaluate the efficiency of the embedding encryption process in \texttt{CAPRISE}. Figure~\ref{fig:efficiency} presents the encryption throughput across varying embedding sizes. We observe that as the embedding size increases, the encryption throughput, measured in embeddings per second, decreases noticeably. While the overall embedding-level throughput decreases with increasing embedding size, Figure~\ref{fig:rate_encryption} shows that the encryption throughput at the element level oscillates but generally remains high when embedding size range from 1536 to 3072. This indicates that the per-element encryption process becomes more efficient for higher-dimensional embeddings.

Specifically, in Phase 1, which involves client-side preprocessing, we evaluate the efficiency of \texttt{CAPRISE} by comparing its runtime to that of pure embedding computation. Empirically, generating embeddings for a batch of 128 queries takes approximately 79.52 ms, while applying \texttt{CAPRISE} encryption to the same batch introduces only an additional 15 ms. This means that the encryption overhead accounts for less than 19\% of the embedding time, indicating that \texttt{CAPRISE} adds minimal computational cost during the embedding phase.

In Phase 2, we evaluate the encryption throughput of \texttt{CAPRISE} against the PHE-based encryption used in RemoteRAG~\cite{cheng2024remoterag}. With an embedding size of 768, RemoteRAG reports an encryption speed of approximately 250–300 vectors per second. In contrast, \texttt{CAPRISE} achieves a significantly higher throughput of 2,339 vectors per second, representing a 9× improvement in efficiency. This substantial gain underscores \texttt{CAPRISE}'s practicality for deployment in latency-sensitive retrieval systems.

\subsubsection{Privacy} \label{sec:exp_privacy}
To evaluate the privacy protection provided by \texttt{CAPRISE}, we examine its effectiveness against the reconstruction-based \textit{Vec2Text} attack~\cite{morris2023text}. Given an input document $x_i$, its embedding $e_i$ can be inverted to reconstruct a text approximation $\hat{x}_i$ using Vec2Text. We measure the semantic similarity between the original text $x_i$ and the reconstructed text $\hat{x}_i$ using metrics: BLEU, Precision, Recall, F1, and ROUGE. 

We compare the reconstruction quality for embeddings before and after applying \texttt{CAPRISE}. Our results, presented in Table~\ref{tab:reconstruction_quality}, show that under the secure parameter setting of $s=3$, $\beta=0.2$, \texttt{CAPRISE} significantly reduces the reconstructability of encrypted embeddings. Notably, the BLEU score drops from over 83.021 on plain embeddings to 12.425 on encrypted ones, indicating the strong privacy guarantees of our approach against embedding inversion attacks.

\begin{table}[t]
\caption{Comparison of Vec2Text Reconstruction w/o and w/ \texttt{CAPRISE}.}
\label{tab:reconstruction_quality}
\centering
\begin{tabular}{l|ccccc}
\toprule
         & BLEU  & Precision & Recall & F1    & ROUGE \\
\midrule
w/o \texttt{Enc} & 83.021 & 0.947      & 0.950   & 0.948  & 0.950  \\
\textbf{\texttt{CAPRISE}} & 12.425 & 0.482      & 0.498   & 0.487  & 0.492  \\
\bottomrule
\end{tabular}
\end{table}

\subsubsection{Defense against Vectors Analysis} \label{sec:exp_accuracy}
To evaluate the privacy strength of our proposed encryption scheme $\mathbf{Enc_{\text{DB}}}$ in \texttt{CAPRISE}, we conduct a vector analysis experiment and compare its robustness against that of \texttt{ADCPE}. The goal is to assess how well each method obscures the underlying embedding structure from inference attacks.

\begin{figure}[h]
    \centering
    \includegraphics[width=.95\linewidth]{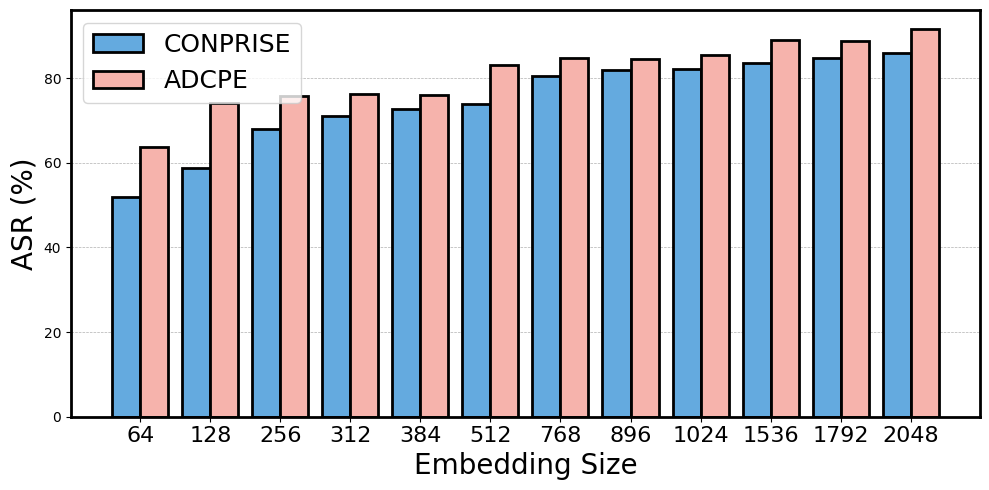}
    \caption{Attack with vectors analysis.}
    \label{fig:exp_asr}
\end{figure}

In this experiment, we randomly select a vector from the database as a test query vector and compute the distances between this query and all other vectors in the dataset. We then encrypt the database using \texttt{CAPRISE} and \texttt{ADCPE}, respectively, and perform the same distance computation using the encrypted vectors. For reference, we also calculate the distances in plaintext space, treating it as the ground truth.

Next, we rank all vectors based on their distance to the query in each setting: plaintext, \texttt{ADCPE} ciphertext, and \texttt{CAPRISE} ciphertext. We then compare how closely the encrypted distance rankings align with the plaintext ranking using Attack Success Rate (ASR). A lower ASR indicates stronger privacy protection, as it implies a larger deviation from the original semantic distance structure. 

As shown in Figure~\ref{fig:exp_asr}, \texttt{CAPRISE} consistently achieves lower ASR than \texttt{ADCPE} across varying embedding sizes, demonstrating its stronger resilience against vector analysis attacks.

\subsubsection{Defense against Query Analysis} \label{sec:exp_dp}
To evaluate the effectiveness of our defense against query analysis attacks, we empirically measure how the perturbation radius $r$ affects the retrieval set expansion. Using an embedding size of $n$ and a database of $m$ encrypted embeddings, we perturb the query embedding with $r$ and compute the required top-$k'$ such that it still contains the top-$k$ results, as described in Equation~\ref{eq:dp_relationship}. As shown in Table~\ref{tab:radius_k_ratio}, a smaller radius results in a lower expansion ratio $k'/k$, meaning the top-$k$ results can be more easily recovered from top-$k'$, which implies weaker privacy. Conversely, a larger radius increases uncertainty and strengthens privacy guarantees. This highlights the inherent trade-off between privacy and retrieval accuracy: stronger privacy requires broader retrieval, which may reduce precision in downstream generation.

\begin{table}[t]
\centering
\caption{Comparison of top-$k'$ and $k'/k$ for different embedding sizes ($n = 768, 1536$) and radius value ($r=0.033 $ and $0.02$) under $m=100,000 $ embeddings.}
\resizebox{\columnwidth}{!}{%
\begin{tabular}{c|cc|cc||cc|cc}
\toprule
\multirow{3}{*}{top-$k$} 
& \multicolumn{4}{c||}{$n = 768$} 
& \multicolumn{4}{c}{$n = 1536$} \\
\cmidrule(lr){2-5} \cmidrule(lr){6-9}
& \multicolumn{2}{c|}{top-$k'$} & \multicolumn{2}{c||}{$k'/k$}
& \multicolumn{2}{c|}{top-$k'$} & \multicolumn{2}{c}{$k'/k$} \\
& $0.033$ & $0.02$ & $0.033$ & $0.02$ 
& $0.033$ & $0.02$ & $0.033$ & $0.02$ \\
\midrule
5  & 258  & 58  & 52  & 12 & 982  & 145 & 196 & 29 \\
10 & 571  & 108 & 57  & 11 & 1492 & 245 & 149 & 25 \\
15 & 673  & 170 & 45  & 11 & 2077 & 388 & 138 & 26 \\
20 & 928  & 203 & 46  & 10 & 2603 & 501 & 130 & 25 \\
\bottomrule
\end{tabular}
}
\label{tab:radius_k_ratio}
\end{table}

\section{Discussions}
\label{sec:discussions}
 In real-world deployment, one key challenge lies in adapting ppRAG across different datasets and embedding models. Since different datasets exhibit distinct feature distributions and embedding geometries, the encryption parameter $\beta$, which controls the distance-preserving property in CAPRISE, cannot be generally fixed. Each embedding model may require a dataset-specific calibration of $\beta$ to maintain both retrieval fidelity and privacy strength. Similarly, during the query phase, the differential privacy mechanism introduces perturbation within a configurable range that determines the trade-off between privacy and retrieval accuracy. A larger perturbation radius enhances privacy but increases retrieval noise and computational cost, whereas a smaller radius preserves precision at the expense of weaker protection. Future research can focus on systematically studying this privacy-utility trade-off, developing adaptive parameter selection strategies that dynamically adjust $\beta$ and the perturbation range according to dataset characteristics, query privacy, and retrieval requirements.

\section{Conclusion}
In this work, we propose \texttt{ppRAG} to protect user data in untrusted environments. By introducing \texttt{CAPRISE}, ppRAG enables similarity-based retrieval over encrypted embeddings while preventing reconstruction-based and query analysis attacks. Our system further incorporates differential privacy to obfuscate query embeddings and standard encryption to secure retrieved documents, achieving strong end-to-end privacy guarantees. Experimental results demonstrate that ppRAG not only offers robust defenses against vec2text leakage, query analysis, and vector analysis but also maintains high retrieval utility and processing throughput. This work bridges the gap between functional similarity search and rigorous privacy, paving the way for the secure and trustworthy deployment of RAG systems in real-world applications.

\section*{Acknowledgment}


This research is supported by the National Research Foundation, Singapore and Infocomm Media Development Authority under its Trust Tech Funding Initiative. Any opinions, findings and conclusions or recommendations expressed in this material are those of the author(s) and do not reflect the views of National Research Foundation, Singapore and Infocomm Media Development Authority.


\bibliographystyle{IEEEtran}
\bibliography{mybibliography}

@inproceedings{morris2023text,
  title={Text Embeddings Reveal (Almost) As Much As Text},
  author={Morris, John and Kuleshov, Volodymyr and Shmatikov, Vitaly and Rush, Alexander M},
  booktitle={Proceedings of the 2023 Conference on Empirical Methods in Natural Language Processing},
  pages={12448--12460},
  year={2023}
}

@article{brown2020language,
  title={Language models are few-shot learners},
  author={Brown, Tom and Mann, Benjamin and Ryder, Nick and Subbiah, Melanie and Kaplan, Jared D and Dhariwal, Prafulla and Neelakantan, Arvind and Shyam, Pranav and Sastry, Girish and Askell, Amanda and others},
  journal={Advances in neural information processing systems},
  volume={33},
  pages={1877--1901},
  year={2020}
}

@article{ji2023survey,
  title={Survey of hallucination in natural language generation},
  author={Ji, Ziwei and Lee, Nayeon and Frieske, Rita and Yu, Tiezheng and Su, Dan and Xu, Yan and Ishii, Etsuko and Bang, Ye Jin and Madotto, Andrea and Fung, Pascale},
  journal={ACM Computing Surveys},
  volume={55},
  number={12},
  pages={1--38},
  year={2023},
  publisher={ACM New York, NY}
}

@article{lewis2020retrieval,
  title={Retrieval-augmented generation for knowledge-intensive nlp tasks},
  author={Lewis, Patrick and Perez, Ethan and Piktus, Aleksandra and Petroni, Fabio and Karpukhin, Vladimir and Goyal, Naman and K{\"u}ttler, Heinrich and Lewis, Mike and Yih, Wen-tau and Rockt{\"a}schel, Tim and others},
  journal={Advances in Neural Information Processing Systems},
  volume={33},
  pages={9459--9474},
  year={2020}
}

@inproceedings{izacard2021leveraging,
  title={Leveraging Passage Retrieval with Generative Models for Open Domain Question Answering},
  author={Izacard, Gautier and Grave, {\'E}douard},
  booktitle={Proceedings of the 16th Conference of the European Chapter of the Association for Computational Linguistics: Main Volume},
  pages={874--880},
  year={2021}
}

@inproceedings{curtmola2006searchable,
  title={Searchable symmetric encryption: improved definitions and efficient constructions},
  author={Curtmola, Reza and Garay, Juan and Kamara, Seny and Ostrovsky, Rafail},
  booktitle={Proceedings of the 13th ACM conference on Computer and communications security},
  pages={79--88},
  year={2006}
}

@article{cheng2024remoterag,
  title={RemoteRAG: A Privacy-Preserving LLM Cloud RAG Service},
  author={Cheng, Yihang and Zhang, Lan and Wang, Junyang and Yuan, Mu and Yao, Yunhao},
  journal={arXiv preprint arXiv:2412.12775},
  year={2024}
}

@article{koga2024privacy,
  title={Privacy-Preserving Retrieval Augmented Generation with Differential Privacy},
  author={Koga, Tatsuki and Wu, Ruihan and Chaudhuri, Kamalika},
  journal={arXiv preprint arXiv:2412.04697},
  year={2024}
}

@article{gao2023retrieval,
  title={Retrieval-augmented generation for large language models: A survey},
  author={Gao, Yunfan and Xiong, Yun and Gao, Xinyu and Jia, Kangxiang and Pan, Jinliu and Bi, Yuxi and Dai, Yi and Sun, Jiawei and Wang, Haofen},
  journal={arXiv preprint arXiv:2312.10997},
  year={2023}
}

@article{wu2024retrieval,
  title={Retrieval-augmented generation for natural language processing: A survey},
  author={Wu, Shangyu and Xiong, Ying and Cui, Yufei and Wu, Haolun and Chen, Can and Yuan, Ye and Huang, Lianming and Liu, Xue and Kuo, Tei-Wei and Guan, Nan and others},
  journal={arXiv preprint arXiv:2407.13193},
  year={2024}
}

@inproceedings{chen2024benchmarking,
  title={Benchmarking large language models in retrieval-augmented generation},
  author={Chen, Jiawei and Lin, Hongyu and Han, Xianpei and Sun, Le},
  booktitle={Proceedings of the AAAI Conference on Artificial Intelligence},
  volume={38},
  number={16},
  pages={17754--17762},
  year={2024}
}

@article{li2022survey,
  title={A survey on retrieval-augmented text generation},
  author={Li, Huayang and Su, Yixuan and Cai, Deng and Wang, Yan and Liu, Lemao},
  journal={arXiv preprint arXiv:2202.01110},
  year={2022}
}

@inproceedings{zhuang2024understanding,
  title={Understanding and mitigating the threat of vec2text to dense retrieval systems},
  author={Zhuang, Shengyao and Koopman, Bevan and Chu, Xiaoran and Zuccon, Guido},
  booktitle={Proceedings of the 2024 Annual International ACM SIGIR Conference on Research and Development in Information Retrieval in the Asia Pacific Region},
  pages={259--268},
  year={2024}
}

@inproceedings{fuchsbauer2022approximate,
  title={Approximate distance-comparison-preserving symmetric encryption},
  author={Fuchsbauer, Georg and Ghosal, Riddhi and Hauke, Nathan and O’Neill, Adam},
  booktitle={International Conference on Security and Cryptography for Networks},
  pages={117--144},
  year={2022},
  organization={Springer}
}

@inproceedings{ma2023query,
  title={Query Rewriting in Retrieval-Augmented Large Language Models},
  author={Ma, Xinbei and Gong, Yeyun and He, Pengcheng and Duan, Nan and others},
  booktitle={The 2023 Conference on Empirical Methods in Natural Language Processing}
}

@inproceedings{peng2024large,
  title={Large language model based long-tail query rewriting in taobao search},
  author={Peng, Wenjun and Li, Guiyang and Jiang, Yue and Wang, Zilong and Ou, Dan and Zeng, Xiaoyi and Xu, Derong and Xu, Tong and Chen, Enhong},
  booktitle={Companion Proceedings of the ACM on Web Conference 2024},
  pages={20--28},
  year={2024}
}

@inproceedings{gao2023precise,
  title={Precise Zero-Shot Dense Retrieval without Relevance Labels},
  author={Gao, Luyu and Ma, Xueguang and Lin, Jimmy and Callan, Jamie},
  booktitle={Proceedings of the 61st Annual Meeting of the Association for Computational Linguistics (Volume 1: Long Papers)},
  pages={1762--1777},
  year={2023}
}

@article{wang2024unims,
  title={Unims-rag: A unified multi-source retrieval-augmented generation for personalized dialogue systems},
  author={Wang, Hongru and Huang, Wenyu and Deng, Yang and Wang, Rui and Wang, Zezhong and Wang, Yufei and Mi, Fei and Pan, Jeff Z and Wong, Kam-Fai},
  journal={arXiv preprint arXiv:2401.13256},
  year={2024}
}

@inproceedings{sundfa,
  title={DFA-RAG: Conversational Semantic Router for Large Language Model with Definite Finite Automaton},
  author={Sun, Yiyou and Hu, Junjie and Cheng, Wei and Chen, Haifeng},
  booktitle={Forty-first International Conference on Machine Learning},
  year={2024}
}

@inproceedings{asaiself,
  title={Self-RAG: Learning to Retrieve, Generate, and Critique through Self-Reflection},
  author={Asai, Akari and Wu, Zeqiu and Wang, Yizhong and Sil, Avirup and Hajishirzi, Hannaneh},
  booktitle={The Twelfth International Conference on Learning Representations},
  year={2024}
}

@inproceedings{wang2023self,
  title={Self-Knowledge Guided Retrieval Augmentation for Large Language Models},
  author={Wang, Yile and Li, Peng and Sun, Maosong and Liu, Yang},
  booktitle={The 2023 Conference on Empirical Methods in Natural Language Processing}
}

@inproceedings{linra,
  title={RA-DIT: Retrieval-Augmented Dual Instruction Tuning},
  author={Lin, Xi Victoria and Chen, Xilun and Chen, Mingda and Shi, Weijia and Lomeli, Maria and James, Richard and Rodriguez, Pedro and Kahn, Jacob and Szilvasy, Gergely and Lewis, Mike and others},
  booktitle={The Twelfth International Conference on Learning Representations},
  year={2024}
}

@inproceedings{wanginstructretro,
  title={InstructRetro: Instruction Tuning post Retrieval-Augmented Pretraining},
  author={Wang, Boxin and Ping, Wei and McAfee, Lawrence and Xu, Peng and Li, Bo and Shoeybi, Mohammad and Catanzaro, Bryan},
  booktitle={Forty-first International Conference on Machine Learning},
  year={2024}
}

@article{nguyen2016ms,
  title={MS MARCO: A Human Generated MAchine Reading COmprehension Dataset},
  author={Nguyen, Tri and Rosenberg, Mir and Song, Xia and Gao, Jianfeng and Tiwary, Saurabh and Majumder, Rangan and Deng, Li},
  journal={choice},
  volume={2640},
  pages={660},
  year={2016}
}

@inproceedings{ni2022large,
  title={Large Dual Encoders Are Generalizable Retrievers},
  author={Ni, Jianmo and Qu, Chen and Lu, Jing and Dai, Zhuyun and Abrego, Gustavo Hernandez and Ma, Ji and Zhao, Vincent and Luan, Yi and Hall, Keith and Chang, Ming-Wei and others},
  booktitle={Proceedings of the 2022 Conference on Empirical Methods in Natural Language Processing},
  pages={9844--9855},
  year={2022}
}

\end{document}